\documentclass{article}
\usepackage{nips07submit_e,times}
\usepackage[dvips]{graphicx}
\usepackage{amssymb,amsmath,epsfig}
\usepackage{subfig}
\usepackage{color}

\newcommand{\RR}{I\!\!R} 
\newcommand{\bigO}{\mathcal{O}}
\renewcommand{\vec}[1]{\boldsymbol{#1}} 
\newcommand{\Lone}{$l^1$}
\newcommand{\mat}[1]{\textbf{\em #1}} 

\newenvironment{opsomming}%
{\begin{list}{\em $\bullet$}{\partopsep 0ex  %
                             \parskip 0.5ex   %
                             \topsep 0.0ex   %
                             \partopsep 0mm  %
                             \itemsep 0.1ex  %
                             \parsep 0mm     %
                             \itemindent 0mm %
                             \labelwidth 1em %
                             \labelsep   0.5em %
                             \leftmargin 1.5em}}%
{\end{list} \vspace{0.5ex}}

\title{TR01: Time-continuous Sparse Imputation}

\author{
Jort Gemmeke \\
Department of Language and Speech\\
Radboud University Nijmegen\\
The Netherlands \\
\texttt{J.Gemmeke@let.ru.nl} \\
\And
Bert Cranen \\
Department of Language and Speech\\
Radboud University Nijmegen\\
The Netherlands \\
\texttt{B.Cranen@let.ru.nl} \\
}

\begin{document}

\maketitle

\begin{abstract}
An effective way to increase the noise robustness of automatic speech recognition  is to label noisy speech features as either reliable or unreliable (missing) prior to decoding, and to replace the missing ones by clean speech estimates. 
We present a novel method to obtain such clean speech estimates. Unlike previous imputation frameworks which work on a frame-by-frame basis, our method focuses on exploiting information from a large time-context. Using a sliding window approach, denoised speech representations are constructed using a sparse representation of the reliable features in an overcomplete basis of fixed-length exemplar fragments. We demonstrate the potential of our approach with experiments on the 
{\sc aurora-2} connected digit database.
\end{abstract}

\section{Introduction}

Automatic speech recognition (ASR) performance degrades substantially when speech is corrupted by background noise that was not seen during training. Missing Data Techniques (MDTs)~\cite{Raj1998, Cooke2001a} provide a powerful way to mitigate the impact of both stationary and non-stationary noise for a wide range of Signal-to-Noise (SNR) ratios. The general idea behind MDT is that it is possible to estimate $-$prior to decoding$-$  which spectro-temporal elements of the acoustic representations are reliable (i.e., dominated by speech) and which are unreliable (i.e., dominated by background noise). These reliability estimates, referred to as a \emph{spectrographic mask}, are used to treat reliable and unreliable features differently. The mask information can for instance be used to replace the unreliable features by clean speech estimates~(e.g., \cite{Raj2000,Van2004b,Josifovski1999}) which is called \emph{imputation}.
\par
Although, admittedly, impressive gains in recognition accuracy  have been achieved using MDTs, at SNRs $\leq 0$\,dB the performance is often too poor for practical applications. A possible explanation for the problems at low SNRs is the fact that most missing data imputation methods work on a frame-by-frame basis (i.e. strictly local in time). 
However, at SNRs $\leq 0$ dB a substantial number of frames may contain few, if any, reliable features. Therefore, there is an increased risk that individual frames do not contain sufficient information for successful imputation. 
\par
In \cite{Gemmeke2008}, we showed that this data scarcity problem at very low SNRs  can be solved by a missing data imputation method that uses a time window which is (much) wider than a single frame. This allows a better exploitation of the redundancy of the speech signal. The technique, \emph{sparse imputation}, works by finding a sparse representation of the reliable features of an unknown word in an overcomplete basis of noise-free example words. The projection of these sparse representations in the basis is then used to provide clean speech estimates to replace the unreliable features. Since the imputation framework introduced in \cite{Gemmeke2008} represents each word by a fixed-length vector, its applicability is limited to situations where the word boundaries are known beforehand, such as in isolated word recognition.
\par
In the current paper we extend sparse imputation for use in continuous speech recognition. Rather than imputing whole words using a basis of exemplar words, we impute fixed-length sliding time windows using a basis with examples of fixed-length fragments of clean speech.
Our goal is to establish to what extent this approach leads to better recognition accuracies  at SNRs $\leq 0$ dB compared to conventional ASR methods. The technique might bring practical applications within reach that  are substantially less vulnerable to noise. We evaluate our novel approach by comparing its performance with that of a state-of-the-art frame-based imputation approach, using the {\sc aurora-2} continuous digit recognition task \cite{Hirsch2000}. First, we give an upper bound on the performance of both techniques by using `oracle' masks~\footnote{Oracle masks are masks in which reliability decisions are based on a priori knowledge, not available in practical settings, about the extent to which each time-frequency cell is dominated by either noise or speech.}. Then we proceed to using an estimated harmonicity mask \cite{Van2004a}.
\par
The rest of the paper is organized as follows. In Section \ref{sec:MD-ASR} we briefly describe MDT. In Section \ref{sec:sparseimputation} we introduce the sparse imputation framework. In Section \ref{sec:contimputation} we extend this framework for use in continuous ASR. In Section \ref{sec:experiments} we compare recognition accuracies with the baseline decoder and we give our conclusions in Section \ref{sec:conclusions}. We conclude with a description of future work.

\section{Missing Data Techniques}\label{sec:MD-ASR}
In ASR, speech representations are typically based on some spectro-temporal distribution of acoustic power, called a spectrogram.
In noise-free conditions, the value of each element in this two-dimensional matrix is determined by the speech signal only. In noisy conditions, the acoustic power in each cell may also (in part) be due to background noise. 
Assuming the noise is additive the spectrogram of noisy speech, denoted by $\mat{Y}$, can be described as the sum of the individual spectrograms of clean speech $\mat{S}$ and noise $\mat{N}$, i.e., $\mat{Y}=\mat{S}+\mat{N}$. 
Elements of $\mat{Y}$ that predominantly contain speech or noise energy are distinguished by introducing a spectrographic mask. With all spectrograms represented as $K \times T$ dimensional matrices ($K$ being the number of frequency bands and $T$ the number of time frames), a mask is defined as an equally sized matrix. Its elements are either $1$, meaning the corresponding cell of $\mat{Y}$ is dominated by speech (`reliable') or $0$, meaning it is dominated by noise (`unreliable' c.q. `missing'). 
Thus, we write:

\begin{equation}\label{eq:maskdef}
M(k,t) =
\left\{
\begin{array}{ll}
\mbox{1 $\stackrel{def}{=}$ reliable}   & \mbox{if}~S(k,t) > N(k,t)\\
\mbox{0 $\stackrel{def}{=}$ unreliable} & \mbox{otherwise}
\end{array}
\right.
\end{equation}

\noindent with frequency band $k$ ($1 \leq k \leq K$)  and time frame $t$ ($1 \leq t \leq T$). Then, if the power spectrum of the noisy speech is represented on a log-compressed scale, we may write for reliable features:

\begin{equation}
\log[Y(k,t)] = \log[S(k,t)\cdot(1+N(k,t)/S(k,t))] \approx \log[S(k,t)]
\end{equation}

\noindent 
In other words, under the assumption of additive background noise, reliable noisy speech coefficients can be used directly as  estimates of the clean speech features.
\par
In experiments with artificially added noise, the mask can be computed using knowledge about the corrupting noise and the clean speech signal, the so-called oracle masks. In realistic situations, however, the masks must be estimated. Many different estimation techniques have been proposed, such as SNR based estimators \cite{Vizinho1999}, methods that focus on speech characteristics, e.g. harmonicity based SNR estimation \cite{Van2004a} and mask estimation by means of Bayesian classifiers \cite{Kim2006}. We refer the reader to \cite{Cerisara2007} and the references therein for a more complete overview of mask estimation techniques. 
In Section~ \ref{sec:experiments} we will use one of these masks (i.c. the harmonicity mask \cite{Van2004a}) to illustrate the properties of our method in combination with an estimated mask. 
\par
Techniques for ASR with missing data can be divided into \emph{imputation} and \emph{marginalization}. With marginalization \cite{Cooke2001a} missing values are ignored during the decoding by integrating over their possible ranges. 
With imputation \cite{Raj2000} missing features are replaced by estimates (expected values extracted from the training set). In this paper we will only consider imputation. Imputation may be viewed as a data cleaning technique, enabling the use of conventional ASR systems that perform recognition as if all features were reliable. Imputation techniques may also be integrated in an ASR engine as illustrated by a successful approach called \emph{conditioned imputation} \cite{Van2004b}. The latter approach, which we will use in Section~\ref{sec:experiments} to compare our new method against, makes the clean speech estimates dependent on the hypothesized state of the hidden Markov model. Furthermore, it imposes the additional constraint that the power of the clean speech estimates must not exceed the observed noisy speech power.

\section{Imputation using sparse representations}\label{sec:sparseimputation}

\subsection{Sparse representation of speech signals}\label{subsec:sparseimputation:sparserep}
We express the $K \times T$ spectrogram matrix of noisy speech $\mat{Y}$ as a single vector $\vec{y}$ of dimension $D = K \cdot T$ by concatenating $T$ subsequent time frames. For the moment, we assume $T$ to be fixed, which in practice means we have to time-normalize all utterances we want to process. As in \cite{Gemmeke2008}, we consider $\vec{y}$ to be a non-negative linear combination of exemplar spectrograms $\vec{a}_{n}$, where $n$ denotes a specific exemplar $(1\leq n \leq N_A)$ in the set of $N_A$ available exemplars. We write:

\begin{equation}
 \vec{y} = \sum_{n=1}^{N_A}x_{n} \vec{a_{n}} = \mat{A}\vec{x}
\end{equation}

\noindent with weights $x_{n} \geq 0 \in \RR$, $\vec{x}$ an $N_A$-dimensional weight vector, and $\mat{A}=\left(\vec{a_{1}}~~\vec{a_{2}} \ldots \vec{a_{N-1}}~~\vec{a_{N}} \\[1.05ex] \right)$ a matrix with dimensionality $D \times N$.
\par
Typically, the number of exemplar spectrograms will be much larger than the dimensionality of the acoustic representation ( $N_A \gg D$). Therefore, the system of linear equations has no unique solution. Research in the field of \emph{compressed sensing} \cite{Donoho2006a,Candes2006b} has shown however that if $\vec{x}$ is \emph{sparse}, $\vec{x}$ can be determined \emph{exactly} by solving:
\begin{equation}\label{minl0}
 \min_{\vec{x}}\{\,\|\vec{x}\|_0\,\} \mbox{ subject to } \vec{y} = \mat{A}\vec{x}
\end{equation}
with $\| . \|_0$ the $l^0$ zero norm (i.e., the number of nonzero elements).

\subsection{\Lone~minimization}\label{sec:method:l1}
The combinatorial problem in Eq.~\ref{minl0} is NP-hard and therefore cannot be solved in practical applications. However, it has been proven that, with mild conditions on the sparsity of $\vec{x}$ and the structure of $A$, $\vec{x}$ can be determined \cite{Donoho2006b} by solving:

\begin{equation}\label{minl1}
 \min_{\vec{x}}\{\,\|\vec{x}\|_1\,\} \mbox{ subject to } \vec{y} = \mat{A}\vec{x}
\end{equation}
This convex minimization problem can be cast as a least squares problem with an \Lone~penalty:

\begin{equation}\label{minl2l1}
  \min_{\vec{x}}\{\,\|\mat{A}\vec{x} -\vec{y}\|_2 + \lambda\|\vec{x}\|_1\,\}
\end{equation}
with a regularization parameter $\lambda$ and a non-negativity constraint on $\vec{x}$. If $\vec{x}$, with sparsity $f=\|\vec{x}\|_0$, is very sparse, Eq.~\ref{minl2l1} can be solved efficiently in $\bigO(f^3 + N_A)$ time using homotopy methods \cite{Efron2004}.\\

\subsection{Sparse imputation}\label{subsec:sparseimput}

To distinguish between reliable and unreliable features in $\vec{y}$ we do not solve Eq.~{\ref{minl2l1} directly, but carry out a \emph{weighted} norm minimization instead:

\begin{equation}\label{minl1weight}
  \min_{\vec{x}}\{\|\mat{W}\mat{A}\vec{x} -\mat{W}\vec{y}\|_2 + \lambda\|\vec{x}\|_1\}
\end{equation}

\noindent with $\mat{W}$ a diagonal matrix of which the elements are determined directly by the binary missing data mask $\mat{M}$ and are either $0$ or $1$. By concatenating subsequent time frames of $\mat{M}$, similarly as we did for the spectrogram $\mat{Y}$, we construct a vector $\vec{m}$ to represent the weights on the diagonal of $\mat{W}$:~$\mbox{diag}(W)=\vec{m}$. Thus, we effectively use $\mat{W}$ as a row selector picking only those rows of $\mat{A}$ and $\vec{y}$ that are assumed to contain reliable data.

As suggested in \cite{Zhang2006} it is possible to use the sparse representation $\vec{x}$ obtained from solving Eq.~\ref{minl1weight} to estimate the missing values of $\vec{y}$ by reconstruction:

\begin{equation}\label{recony}
  \vec{\hat{y}} = \mat{A}\vec{x}
\end{equation}

$\vec{\hat{y}}$ is obtained by a linear combination of corresponding elements of the basis vectors, the weights of which were determined using only reliable data. Hence, a version of $\vec{\hat{y}}$ that is reshaped into a $K \times T$ matrix can be considered a denoised spectrogram of the underlying speech signal.

\subsection{Theoretical bounds on successful imputation}\label{subsec:thebound}
Obviously, no restoration is possible if $\vec{y}$ does not contain any reliable coefficients at all. In practice, a minimum number of reliable coefficients will be required for successful restoration of $\vec{y}$. While theoretical bounds exist (cf. \cite{Donoho2006a,Candes2006b, Zhang2006}) these are not of great practical value because they depend both on the structure of $\mat{W}\mat{A}$ and the sparsity of $\vec{x}$. Unfortunately, $\mat{W}\mat{A}$ changes from utterance to utterance as the environmental noise changes. Furthermore, the bounds are NP-hard to establish. Hence, we always perform sparse imputation except when no reliable features are present at all, thus accepting the risk of a flawed restoration.

\section{Generalization to time-continuous imputation}\label{sec:contimputation}
The approach described in Sec.~\ref{sec:sparseimputation} is suited for speech units and exemplars that can be adequately represented by an equal number of time frames  $T$, e.g. in isolated word recognition \cite{Gemmeke2008}. However, this does not make sense for arbitrary length utterances and can therefore not be applied to continuous speech recognition. In this section we extend the sparse imputation framework for use with speech signals of arbitrary length by using a sliding, fixed-length time window. Robustness against windows with few or no reliable features is provided by using overlapping windows.

\subsection{Time-shifted imputation}

We divide an utterance $\vec{y}$ of $T$ frames in a series of overlapping time-windows of $R$ frames and perform imputation for every individual window with the method described in  Section~\ref{sec:sparseimputation}. As illustrated in Figure~\ref{fig:shift_schema}, imputation of feature values that belong to overlapping windows is done by averaging the imputed feature values in the individual windows. As before, the basis $\mat{A}$ is formed by $N_A$ exemplar vectors, which are reshaped versions of spectrograms (spanning $R$ frames).  With the spectrogram dimensions being $K \times R$, the vectors have size $L = K \cdot R$, and the dimensions of $\mat{A}$ are $L \times N_A$.
\par
The number of windows $I$ needed for processing the entire speech signal $\vec{y}$ of dimension $D = K \cdot T$ is given by $I = (\mbox{ceil}(D - L)/\Delta)+1$, with $\Delta$ the window shift expressed as  the number of rows in $\vec{y}$ over which the window is shifted. $\Delta$ is a multiple of $K$ because $\vec{y}$ is a vector of concatenated frames, each with $K$ coefficients.
We denote the row indices of $\mat{W}$ and $\vec{y}$ that correspond to the coefficients in the $i^{th}$ window by $\tau$ (with both $i$ and $\tau$ representing natural numbers and $1 \leq i \leq I$ and $i\Delta \leq \tau  \leq i\Delta+L$). At the beginning of  the utterance there will be $L$ of such rows; in the final window, this number reduces from $L$ to $D-I \cdot \Delta$~~(cf. Figure~\ref{fig:shift_schema}). For every window we compute a sparse representation $\vec{x}$ as follows:

\begin{equation}\label{minl1maskcont}
  \min_{\vec{x}}\{\, \|\mat{W}_{\tau}\mat{A}\vec{x} -\mat{W}_{\tau}\vec{y}_{\tau}\|_2 + \lambda\|\vec{x}\|_1\,\}\\
 \end{equation}

The imputed spectrogram for that window, which we will denote by $\gamma$, is computed as $\gamma=\mat{A}\vec{x}$. The use of overlapping imputation windows results in multiple imputation candidates . As depicted in Fig.~\ref{fig:shift_schema}, we have chosen to compute the final clean speech estimate of the $d^{th}$ component of $\hat{\vec{y}}$, denoted by $\hat{y}_d$, as the average of all imputation candidates resulting from overlapping windows. The number of imputation candidates ranges from $1$ (at the beginning and end of an utterance) to $\mbox{ceil}(L/\Delta)$.
\begin{figure}

\centering
 \centerline{\includegraphics[width=0.9\linewidth]{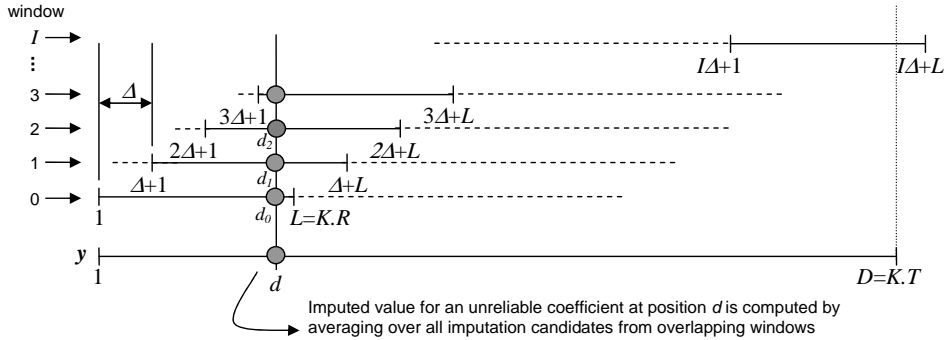}}
\caption{Schematic diagram of imputation using overlapping windows}
\label{fig:shift_schema}
\end{figure}

\section{Experiments}\label{sec:experiments}

To compare the recognition accuracies obtained with the sparse imputation method with those of a conventional, frame based MDT approach, we use a continuous digit recognition task. First, we determine the maximum achievable recognition accuracy for both methods when a priori information is provided about speech and noise in the form of an oracle mask. Second, we study the behaviour of the new imputation method using an estimated mask.
\par
Recognition performance through sparse imputation may be affected by three parameters:  the basis size $N_A$, the window size $R$, and the window shift $\Delta$. In this paper, we keep $N_A$ and $R$ fixed and first investigate how recognition accuracy varies with window shift. In the remaining experiments we study the differences in recognition performance between the sparse imputation method and a conventional frame-based recognizer in more detail using only the best scoring window-shift.

\subsection{Experimental setup}

The speech material used for evaluation is taken from test set `A' of the AURORA-2 corpus \cite{Hirsch2000}. The utterances contain one to seven digits, artificially mixed with four different types of noise, viz. subway, car, babble, exhibition hall. We evaluate recognition accuracy as a function of SNR at the four lowest SNR levels present in the corpus, viz. $10, 5, 0,$ and $-5$\,dB. The results we report are averages over the four noise conditions. The spectrographic representations of the noise $\mat{N}$ and clean speech $\mat{S}$ are available independently. To reduce computation times, we used a random, representative subset of $10\%$ of the utterances (i.e. $\approx 400$ utterances per SNR level).
\par
The exemplar spectrograms in the basis matrix $\mat{A}$ were created by extraction of spectrogram fragments of randomly selected utterances in the clean train set of AURORA-2, using a random offset. The length of the exemplars was chosen $R=35$ frames, which equals the mean number of frames of a single digit \cite{Gemmeke2008}. Thus, the exemplars typically represent sequences of parts of digits. A pilot study with basis sizes ranging from $N_A=4000$ to $N_A=14000$ revealed that recognition accuracy did not increase with $N_A>8000$. We therefore use a basis size $N_A=8000$ throughout this paper. The window shifts experimented with are  $1, 5, 10, 15, 20, 25, 30,$ and $35$ frames.
\par
For the baseline system, we used the state-of-the-art missing data recognition system described in \cite{Van2004b,Van2006}. Acoustic feature vectors consisted of Mel frequency log power spectra ($K=23$ bands). Unreliable features are replaced by estimated values using maximum likelihood per Gaussian based imputation \cite{Van2004b}. The acoustic representations obtained with our sparse imputation method were processed by the baseline system using a spectrographic mask that considers every time-frequency cell as reliable (thus performing no additional missing data imputation).
\par
We use two different masks to describe the reliability of time-frequency cells: 1) an oracle mask and 2) an estimated mask in the form of a harmonicity mask \cite{Van2004a}. The imputation method was implemented in MATLAB. The \Lone~minimization was carried out using the {\tt SolveLasso} solver implemented as part of the {\tt SparseLab} toolbox which can be obtained from www.sparselab.stanford.edu.

\subsection{Results and discussion}
\begin{figure}[!t]
  \centering
\subfloat[][]{\label{fig:shiftsizeoracle}\includegraphics[width=0.45\linewidth]{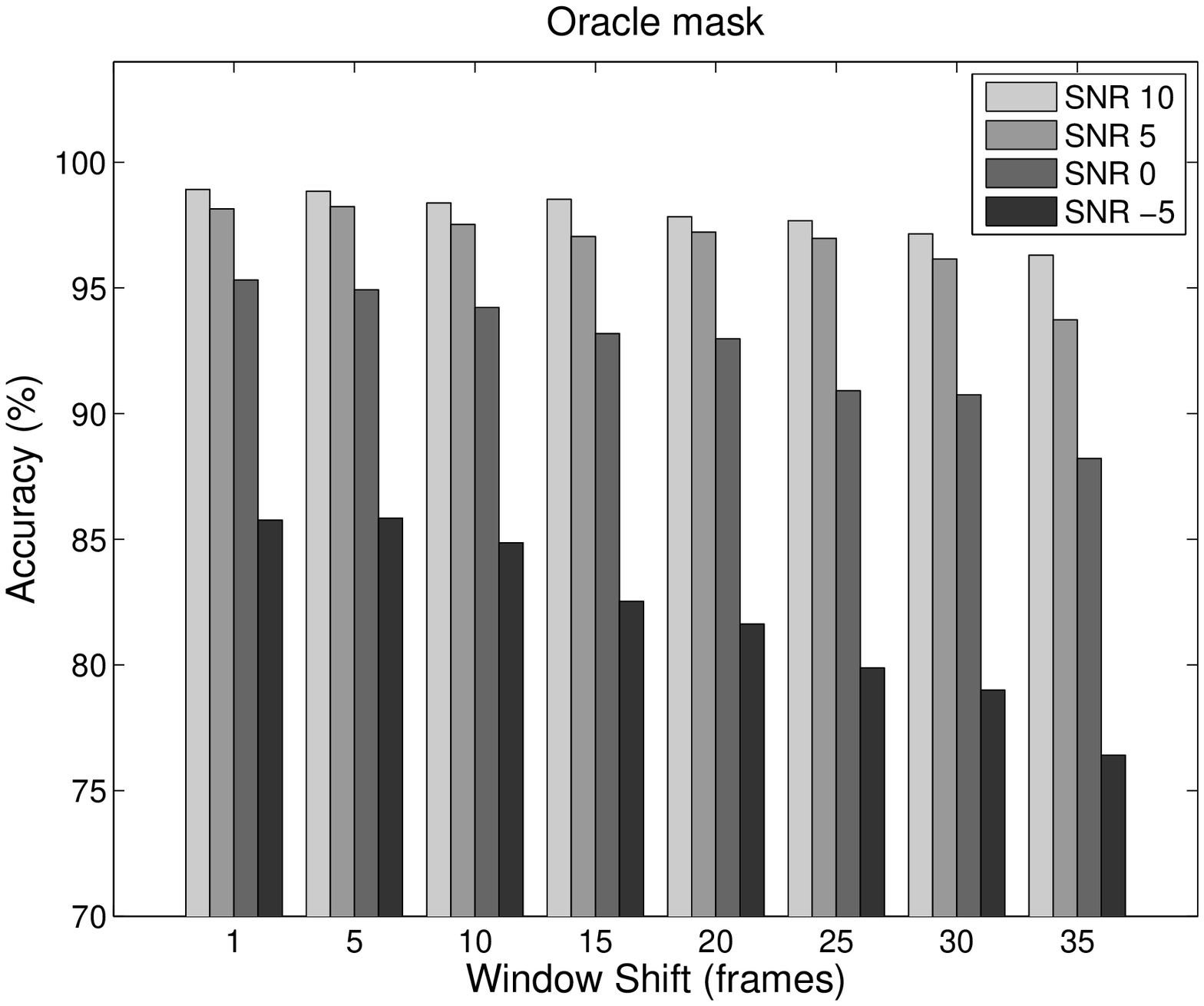}}
\qquad
\subfloat[][]{\label{fig:shiftsizeharm}\includegraphics[width=0.45\linewidth]{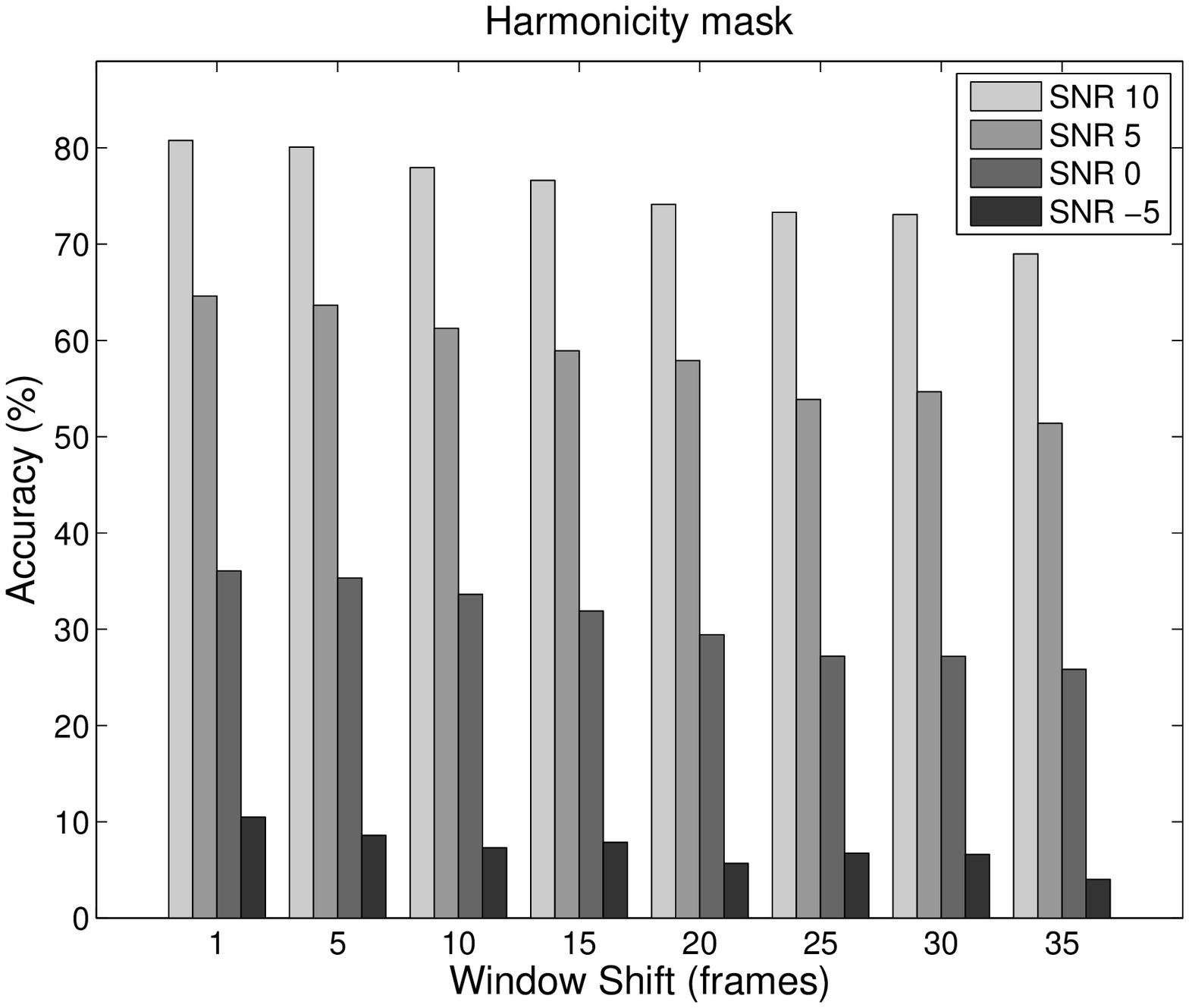}}
  \caption{Word recognition accuracy as a function of window shift. The left pane shows results for the oracle mask and the right pane for the harmonicity mask. Window shift is expressed in frames.}
\end{figure}

\subsubsection{Speech recognition accuracy as a function of window shift}
Figures~\ref{fig:shiftsizeoracle} (for the oracle mask) and \ref{fig:shiftsizeharm} (for the harmonicity mask) show recognition accuracy as a function of the window shift in frames. Both figures show that recognition accuracy steadily decreases as the window shift increases. Moreover, for the oracle mask the performance at low SNRs decreases faster for larger window shifts. This is most likely due to the number of windows with few or no reliable features: the larger the window shift, the fewer overlapping windows there are. As a consequence, the number of windows containing insufficient reliable features for succesful data restoration increases. The results show that the best results are obtained using a window shift of one frame, corresponding to $\Delta=K$. This shift will be used in the remainder of this paper.

\subsubsection{Comparison with baseline decoder: oracle mask}
When used with an oracle mask, the sparse imputation achieves much higher recognition accuracies than the baseline (cf. filled circles in Figure~\ref{fig:digitrecog}). In contrast with the $56\%$ recognition accuracy obtained by the baseline decoder at SNR$=-5$\,dB, $86\%$ is a major improvement. While one should be aware that these results constitute an upper bound on recognition accuracy, it is promising to observe that  the unreliable features can be reconstructed so well, even at very low SNRs, provided the reliable features can be identified correctly.
\par
The improvement of $30\%$ over the baseline decoder is similar to the improvement of $31\%$ reported in \cite{Gemmeke2008} for isolated digit recognition. This corroborates the potential of sparse imputation for ASR: To our knowledge this is the first missing data technique that successfully exploits information from larger time-windows and can be combined with conventional continuous speech decoding. 

\subsubsection{Comparison of accuracies using oracle versus harmonicity mask}
The results obtained with the estimated harmonicity mask are depicted by diamonds in Figure~\ref{fig:digitrecog}. Clearly, the recognition accuracies are much lower than with the oracle mask, suggesting that the harmonicity mask does not succeed in identifying all reliable coefficients as such. Indeed, Fig.~\ref{fig:underandfalse} shows that the percentage of features that is labeled reliable, is substantially lower than in the oracle mask. Yet, the lower recognition accuracies cannot solely be attributed to the reduced number of reliable features. For example, consider the recognition accuracy with sparse imputation for the harmonicity mask at SNR $=5$\,dB. The number of reliable features is roughly equal to that of the oracle mask at SNR $=-5$\,dB, while the recognition accuracy is much lower (65\% vs. 86\%), indicating that the reliable features of the harmonicity mask lack crucial information. However, the fact that the sparse imputation accuracies are lower than those of the baseline also indicates that the current implementation of sparse imputation does not use all information that {\em is} available.

\par
\begin{figure}[!t]
\begin{minipage}[t]{0.48\linewidth}
  \centering
\centerline{\includegraphics[width=1.0\linewidth]{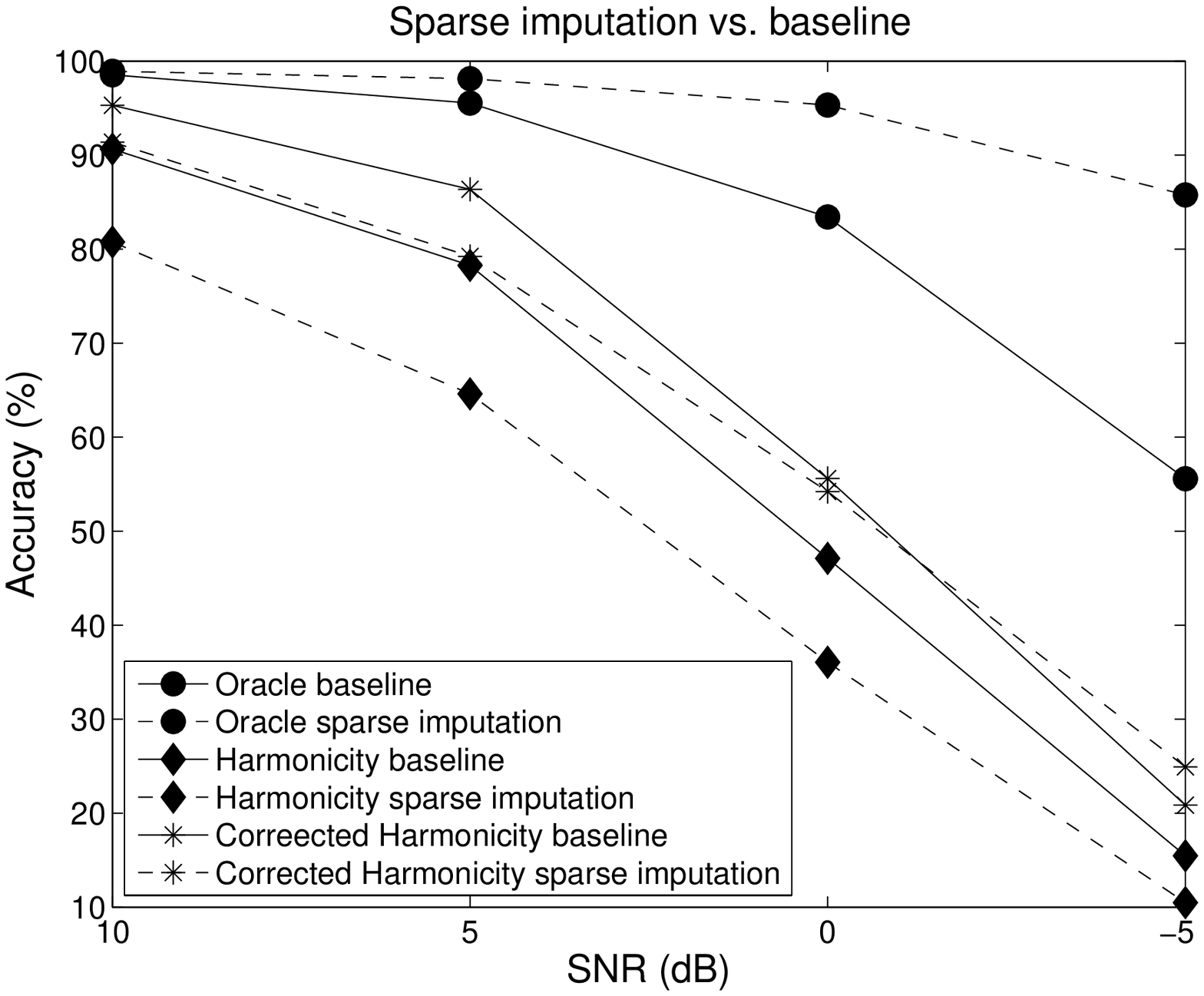}}
  \caption{Word recognition accuracy for both the baseline decoder and the sparse imputation method using the oracle mask, the harmonicity mask and the corrected harmonicity mask respectively. The window shift is one frame.}\label{fig:digitrecog}.
\end{minipage}
\hfill
\begin{minipage}[t]{0.48\linewidth}
  \centering
  \centerline{\includegraphics[width=1.0\linewidth]{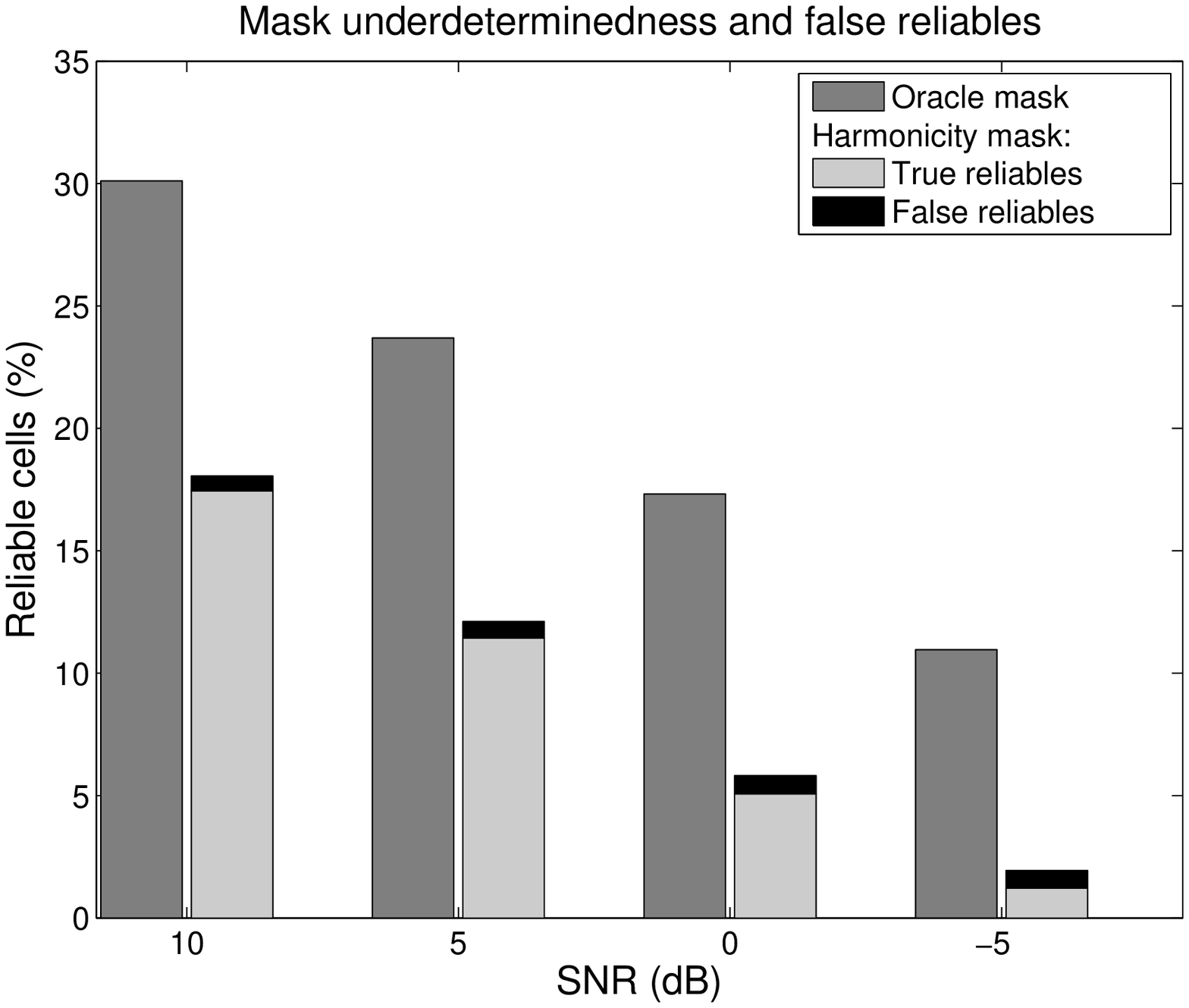}}
  \caption{Percentage of time-frequency cells classified as reliable in the oracle mask and the harmonicity mask. Additionally, the percentage of false reliables in the harmonicity mask is shown.}\label{fig:underandfalse}.
\end{minipage}
\end{figure}

\subsubsection{Comparison with baseline decoder: harmonicity mask}
It is conceivable that the sparse imputation method is more sensitive to \emph{false reliables}, i.e., features labeled reliable by the harmonicity mask while in fact being unreliable (cf. Fig.~\ref{fig:underandfalse}). In order to test this hypothesis we performed recognition using a corrected version of the harmonicity mask without false reliables. The asterisks in Fig.~\ref{fig:digitrecog} illustrate that sparse imputation now performs better than baseline at SNR$=-5$\,dB, comparable at SNR$=0$\,dB and worse at SNRs$>0$\,dB. Also, the overall increase in recognition accuracy is much larger for the sparse imputation framework than for the baseline decoder, confirming the method is indeed more sensitive for false reliables. 
\par
The fact remains, however, that also with the false reliables of the harmonicity mask removed, there still remains a substantial performance gap compared to oracle performance. This indicates that not only the number of reliable features is important for correct imputation, but also their location in the time-frequency plane. Apparently, the extra features labeled reliable by the oracle mask in comparison to the harmonicity mask contain information that is crucial for a correct imputation: The success of finding a sparse representation depends on the exact structure of $\mat{W}\mat{A}$, as described in Section \ref{subsec:thebound}. 
In comparison with the baseline method, the current implementation of the sparse imputation technique too often finds  an incorrect imputation result. 
\par
When using estimated masks, as opposed to oracle masks, apparently more attention is needed for the constraints that 
determine the sparse solution of Eq.~\ref{minl1weight}.
The sparse imputation technique, unlike the baseline decoder, does not take into account that clean speech estimates are bounded by the observation energy. Adding this as an additional constraint to the minimization in Eq.~\ref{minl1weight} 
 might improve the success of finding the correct imputation.
Another way to further constrain the minimization would be to increase the window length $R$. The chosen window length $R~=~35$, the mean length of the digit, implies that many speech examples contain only parts of digits. Larger windows provide both more contextual information and increase the dimensionality of the minimization problem. Informal pilot tests suggest larger window sizes improve performance, but a systematic investigation of this aspect is left as future research.

\section{Conclusions}\label{sec:conclusions}

We introduced a new method for imputation of missing data in continuous speech recognition. It replaces noise-corrupted features in a sliding window by clean speech estimates which are computed using a sparse representation of the reliable features in an overcomplete basis of exemplar speech fragments. 
Imputation results from overlapping windows are combined by averaging. 
\par
The sparse imputation approach was shown to vastly outperform a classical frame-based approach at low SNRs (accuracy of $86\%$ vs. 56\% at SNR$=-5$\,dB) when tested on a continuous digit recognition task and using an oracle mask.
Furthermore, we showed that overlapping windows increase robustness against windows that coincidentally yield a wrong imputation result.
\par
Using estimated masks, we were not able to achieve similar impressive improvements as with oracle masks. 
Clearly, when the reliabity estimates cannot be guaranteed to be correct, the current implementation easily yields erroneous imputations and additional measures are needed to avoid those. We suggested several ways to introduce additional constraints when searching the optimal sparse representation of reliable features. Finding out which of these are the most effective is the subject of future research.

\subsection*{Future work}
\vspace*{-0.5ex}
Future work will focus on refining the framework in several ways to improve performance:
\opsomming
\item Recognition accuracy might benefit from a larger window size since this would provide extra constraints when finding a sparse representation.
 \item Analogously to our baseline, sparse imputation might also profit from bounded imputation. This would require an additional 
 cost function in Eq.~\ref{minl1weight} that enforces $(1-\mat{W})A\vec{x} \leq (1-\mat{W})\vec{y}$.
 \item Research \cite{Barker2000} has shown that it is beneficial to 
 substitute the hard decision in a binary mask by the \emph{probability} that a certain feature is unreliable. The weighting matrix $\mat{W}$ in Eq.~\ref{minl1maskcont} supports the use of such 'fuzzy' masks without further adaptations to the framework.
 \end{list}

\subsubsection*{Acknowledgments}
The research of Jort Gemmeke was carried out in the {MIDAS} project, granted under the Dutch-Flemish STEVIN program. The project partners are the universities of Leuven, Nijmegen and the company Nuance. We aknowledge usefull discussions with Lou Boves.

\subsubsection*{References}
\renewcommand*{\refname}{}
\renewcommand*{\itemsep}{0ex}
\vspace*{-8mm}
\small{
\renewcommand*{\parsep}{0ex}
\bibliographystyle{IEEEtran}
\bibliography{TechReport01}

\begin{thebibliography}{10}
\providecommand{\url}[1]{#1}
\csname url@samestyle\endcsname
\providecommand{\newblock}{\relax}
\providecommand{\bibinfo}[2]{#2}
\providecommand{\BIBentrySTDinterwordspacing}{\spaceskip=0pt\relax}
\providecommand{\BIBentryALTinterwordstretchfactor}{4}
\providecommand{\BIBentryALTinterwordspacing}{\spaceskip=\fontdimen2\font plus
\BIBentryALTinterwordstretchfactor\fontdimen3\font minus
  \fontdimen4\font\relax}
\providecommand{\BIBforeignlanguage}[2]{{%
\expandafter\ifx\csname l@#1\endcsname\relax
\typeout{** WARNING: IEEEtran.bst: No hyphenation pattern has been}%
\typeout{** loaded for the language `#1'. Using the pattern for}%
\typeout{** the default language instead.}%
\else
\language=\csname l@#1\endcsname
\fi
#2}}
\providecommand{\BIBdecl}{\relax}
\BIBdecl

\bibitem{Raj1998}
B.~Raj, R.~Singh, and R.~Stern, ``Inference of missing spectrographic features
  for robust automatic speech recognition,'' in \emph{Proc. International
  Conference on Spoken Language Processing}, 1998, pp. 1491--1494.

\bibitem{Cooke2001a}
M.~Cooke, P.~Green, L.~Josifovksi, and A.~Vizinho, ``Robust automatic speech
  recognition with missing and unreliable acoustic data,'' \emph{Speech
  Communication}, vol.~34, pp. 267--285, 2001.

\bibitem{Raj2000}
B.~Raj, ``Reconstruction of incomplete spectrograms for robust speech
  recognition,'' Ph.D. dissertation, Camegie Mellon University, 2000.

\bibitem{Van2004b}
H.~Van~hamme, ``Prospect features and their application to missing data
  techniques for robust speech recognition,'' in \emph{Proc. INTERSPEECH-2004},
  2004, pp. 101--104.

\bibitem{Josifovski1999}
L.~Josifovski, M.~Cooke, P.~Green, and A.~Vizinho, ``State based imputation of
  missing data for robust speech recognition and speech enhancement,'' in
  \emph{Proc. of Eurospeech}, 1999.

\bibitem{Gemmeke2008}
A.Anonymous, ``Using sparse representations for missing data imputation in
  noise robust speech recognition,'' \emph{To appear in Proc. of EUSIPCO 2008},
  2008.

\bibitem{Hirsch2000}
H.~Hirsch and D.~Pearce, ``The aurora experimental framework for the
  performance evaluation of speech recognition systems under noisy
  conditions,'' in \emph{Proc. of ISCA ASR2000 Workshop, Paris, France}, 2000,
  pp. 181--188.

\bibitem{Van2004a}
H.~Van~hamme, ``Robust speech recognition using cepstral domain missing data
  techniques and noisy masks,'' in \emph{Proc. of IEEE ICASSP}, vol.~1, 2004,
  pp. 213--216.

\bibitem{Vizinho1999}
A.~Vizinho, P.~Green, M.~Cooke, and L.~Josifovski, ``Missing data theory,
  spectral subtraction and signal-to-noise estimation for robust asr: An
  integrated study,'' in \emph{Proc. of Eurospeech}, 1999, pp. 2407--2410.

\bibitem{Kim2006}
W.~Kim and R.~M. Stern, ``Band-independent mask estimation for missing-feature
  reconstruction in the presence of unknown background noise,'' in \emph{Proc.
  of IEEE ICASSP}, 2006.

\bibitem{Cerisara2007}
C.~Cerisara, S.~Demange, and J.-P. Haton, ``On noise masking for automatic
  missing data speech recognition: A survey and discussion,'' \emph{Comput.
  Speech Lang.}, vol.~21, no.~3, pp. 443--457, 2007.

\bibitem{Donoho2006a}
D.~L. Donoho, ``{Compressed sensing},'' \emph{IEEE Transactions on Information
  Theory}, vol.~52, no.~4, pp. 1289--1306, 2006.

\bibitem{Candes2006b}
E.~J. {Candes}, ``{Compressive sampling},'' in \emph{Proc. of the International
  Congress of Mathematicians}, 2006.

\bibitem{Donoho2006b}
D.~L. Donoho, ``For most large underdetermined systems of linear equations the
  minimal l1-norm solution is also the sparsest solution,''
  \emph{Communications on Pure and Applied Mathematics}, vol.~59, no.~6, pp.
  797--829, 2006.

\bibitem{Efron2004}
B.~Efron, T.~Hastie, I.~Johnstone, and R.~Tibshirani, ``Least angle
  regression,'' \emph{Annals of Statistics}, vol.~32, no.~2, pp. 407--499,
  2004.

\bibitem{Zhang2006}
Y.~Zhang, ``When is missing data recoverable?'' \emph{Technical Report}, 2006.

\bibitem{Van2006}
H.~Van~hamme, ``Handling time-derivative features in a missing data framework
  for robust automatic speech recognition,'' in \emph{Proc. of IEEE ICASSP},
  2006.

\bibitem{Barker2000}
J.~Barker, L.~Josifovski, M.~Cooke, and P.~Green, ``Soft decisions in missing
  data techniques for robust automatic speech recognition,'' 2000, pp.
  373--376.

\end{thebibliography}
}

\end{document}